\documentclass[aps,prb,twocolumn,superscriptaddress,showpacs]{revtex4-1}

\usepackage{graphicx}
\usepackage{color}

\begin{document}

\title{Electrochemical synthesis and superconducting phase diagram of Cu$_x$Bi$_2$Se$_3$}

\author{M.~Kriener}
\email[]{mkriener@sanken.osaka-u.ac.jp}
\author{Kouji~Segawa}
\author{Zhi~Ren}
\author{Satoshi~Sasaki}
\author{Shohei~Wada}
\affiliation{Institute of Scientific and Industrial Research, 
Osaka University, Osaka 567-0047, Japan}
\author{Susumu~Kuwabata}
\affiliation{Department of Applied Chemistry, 
Osaka University, Osaka 565-0871, Japan}
\author{Yoichi~Ando}
\email[Corresponding author: ]{y\_ando@sanken.osaka-u.ac.jp}
\affiliation{Institute of Scientific and Industrial Research, 
Osaka University, Osaka 567-0047, Japan}

\date{\today}

\begin{abstract}

The superconducting Cu$_x$Bi$_2$Se$_3$ is an electron-doped
topological insulator and is a prime candidate of the topological
superconductor which still awaits discovery. The
electrochemical intercalation technique for synthesizing Cu$_{x}$Bi$_2$Se$_3$
offers good control of restricting Cu into the van-der-Waals gap and 
yields samples with shielding fractions of up to $\sim 50$\%. We
report essential details of this synthesis technique and present the established
superconducting phase diagram of $T_c$ vs $x$, along with a diagram of
the shielding fraction vs $x$. Intriguingly, those diagrams suggest that
there is a tendency to spontaneously form small islands of optimum
superconductor in this material.

\end{abstract}

\pacs{74.25.Dw; 74.62.-c; 82.45.Vp; 65.40.gk} 


\maketitle

\section{Introduction}

Topological insulators (TIs) are attracting great interest since they
realize a new state of matter; namely, the bulk of such materials is
insulating, but in contrast to conventional band insulators, their bulk
wave functions exhibit a non-trivial $Z_2$ topology leading to gapless
and hence conductive surface states.\cite{K1,MB,Roy} Those topological
surface states are interesting because they exhibit a Dirac-like energy
dispersion (similar to that in graphene) and a helical spin polarization, both of
which hold promise for various energy-saving device
applications.\cite{hasan10a,moore10a,qi10d} Soon after the theoretical
predictions of candidate materials,\cite{K2,SCZ4} a series of promising 
materials were experimentally
discovered to be three-dimensional (3D) TIs,
\cite{H1,Taskin,Matsuda,Shen1,H4,H5,Sato,Kuroda,Shen2,BTS_Rapid,BTS_Hasan,Xiong,BSTS}
stimulating the search for a
superconducting analogue, {\it i.e.}, a topological superconductor,
which is characterized by a full energy gap in the bulk and the
existence of gapless surface Andreev bound
states.\cite{fu08a,schnyder08a,qi09b,qi10b,linder10b,sato10a} Such a
topological superconductor is predicted to be a key to realizing a
fault-tolerant topological quantum computing,\cite{fu08a} and the
discovery of a concrete example of the topological superconductor would
have a large impact on future technology. In this context, there is an
intriguing prediction that when superconductivity is achieved by doping
a topological insulator, it may realize a topological superconducting
state.\cite{fu10a}

Recently, Hor \textit{et al.}\cite{hor10a} reported superconductivity in
the electron-doped TI material Cu$_{x}$Bi$_2$Se$_3$ for $0.1\leq x \leq
0.3$, where the layered TI compound Bi$_2$Se$_3$ was intercalated with
Cu. Their samples showed superconductivity below critical temperatures
$T_{\rm c}\leq 3.8$ K. In a subsequent study on these samples,\cite{wray10a}
 it was further found that the topological surface states
remain intact upon Cu intercalation. Thus, Cu$_x$Bi$_2$Se$_3$ is the
first promising candidate material to be a topological superconductor,
and it is very important to elucidate the nature of 
its superconducting state. However,
the samples prepared by Hor \textit{et al.} showed superconducting
shielding fractions of only less than 20\% and the resistivity never
really disappeared below $T_{\rm c}$.\cite{hor10a,wray10a} Therefore,
some doubts remained about the bulk nature of the superconducting phase in
Cu$_x$Bi$_2$Se$_3$, and preparations of higher-quality samples were
strongly called for.

In our recent experiments,\cite{kriener11a} we succeeded in synthesizing
Cu$_x$Bi$_2$Se$_3$ samples with shielding fractions of up to 50\%
(depending on the Cu concentration) by applying a different sample
preparation method than the standard melt-growth method employed in
Ref.~\onlinecite{hor10a}. In our high-quality samples, besides observing
zero-resistivity, we were able to measure the specific-heat anomaly in
the superconducting phase, which indicated that the superconductivity in
this material is a bulk feature and, moreover, appears to have a full energy
gap,\cite{kriener11a} which is a prerequisite to topological
superconductivity.\cite{schnyder08a} In this paper, 
we describe in detail the newly applied preparation
method of Cu$_{x}$Bi$_2$Se$_3$ to employ electrochemical intercalation,
and present the electronic phase diagram of $T_{\rm c}$ vs Cu
concentration $x$. We also show how the superconducting shielding
fraction of the electrochemically-prepared samples changes with $x$,
which bears an intriguing implication of intrinsically inhomogeneous
superconductivity.

\section{Sample Preparations}

Pristine Bi$_2$Se$_3$ has a layered crystal structure (R$\bar{3}$m, space
group 166) consisting of stacked Se-Bi-Se-Bi-Se quintuple layers. The
rhombohedral [111] direction is denoted as the $c$ axis and the (111)
plane as the $ab$ plane. The neighboring quintuple layers are only
weakly van-der-Waals bonded to each other. Cu is known to enter
Bi$_2$Se$_3$ either as an intercalant in the van-der-Waals gaps or as a
substitutional defect to replace Bi. In the former case Cu$^{+}$ is
formed and acts as a donor, whereas in the latter case it creates two
holes by replacing three Bi 6$p$ electrons by one Cu 4$s$ electron upon
forming a $\sigma$ bond. Therefore Cu acts as an ambipolar dopant for
Bi$_2$Se$_3$.\cite{caywood70a,vasko74a} The melt-growth method employed
by Hor \textit{et al.}\cite{hor10a} cannot avoid the formation of
substitutional Cu defects; in contrast, our electrochemical method has a
distinct advantage to promote only the intercalation of Cu.

\begin{figure}[t]
\includegraphics[width=8.5cm,clip]{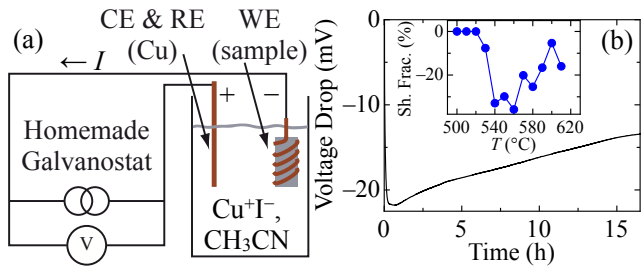}
\caption{(color online)
(a) Sketch of our simplified electrochemical intercalation setup. The
counter electrode (CE) is also used as reference electrode (RE). The
current direction is indicated. (b) Typical time dependence of the
voltage drop measured between working and counter electrode. The inset
summarizes the development of the diamagnetism (plotted in terms of the 
shielding fraction) in a sample with $x$ =
0.31 upon annealing at different temperatures.}
\label{setup}
\end{figure}

For the present experiments, we first grew single crystals of pristine
Bi$_2$Se$_3$ by melting stoichiometric amounts of high-purity elemental
shots of Bi (99.9999\%) and Se (99.999\%) at 850$^{\circ}$C for 96 h in
sealed evacuated quartz glass tubes, followed by a slow cooling to
550$^{\circ}$C over 72 h and annealing at that temperature for 48 h.
Before the electrochemical intercalation procedure, those melt-grown
Bi$_2$Se$_3$ single crystals were cleaved and cut into rectangular
pieces. They were wound by 50 $\mu$m thick, bare Cu wire and acted as
the working electrode (WE). A 0.5-mm thick Cu stick was used both as the
counter (CE) and reference electrode (RE) in our simplified setup
sketched in Fig.~\ref{setup}(a).

Since Hor \textit{et al.}\cite{hor10a} reported that
Cu$_{x}$Bi$_2$Se$_3$ samples are sensitive to air, the pristine
Bi$_2$Se$_3$ samples were transferred into a glove box and the
electrochemical intercalation of Cu was done in an inert atmosphere. To
provide the electrical wiring required for electrochemical intercalation
processes as sketched in Fig.~\ref{setup}(a), our glove box was
specially modified for air-tight electrical connections. For the
intercalation we used a saturated solution of CuI powder (99.99\%) in
acetonitrile CH$_3$CN. A current of 10 $\mu$A was applied to the
electrodes for a suitable time period to give the desired Cu
concentration $x$. The typical time dependence of the voltage drop
between the standard and working electrodes observed for our samples is
shown in Fig.~\ref{setup}(b). Its absolute value was usually between $-10$
and $-30$ mV. The whole process was controlled by a computer which was
also used to determine the total transferred charge. 

After the intercalation had finished, the Cu concentration $x$ in the
sample was determined from its weight change: we measured the weight
before and after the intercalation process with a high-precision balance
with a resolution of 0.1 $\mu$g, and the possible error in $x$ was less
than $\pm$0.01. Note that this direct measurement of the Cu
concentration is expected to be more accurate than most of the
chemical analysis methods.\cite{accuracy} Nevertheless, we have also
employed the inductively-coupled plasma atomic-emission spectroscopy
(ICP-AES) analysis, which is a destructive method but is good at giving
absolute numbers, to confirm that the $x$ values determined from the
weight change is indeed reliable, as will be described later.

For a consistency check of our electrochemical intercalation process, we
have calculated the total charge transferred during the intercalation. The
experimentally determined Faraday number for our samples was usually up
to 15\% smaller than the theoretical Faraday constant, 96485 C/mol. In
this regard, it was reported \cite{caywood70a,vasko74a} that a simple
physical attachment of Cu metal to Bi$_2$Se$_3$ crystal results in a
perceptible diffusion of Cu$^+$ into the van-der-Waals gaps of
Bi$_2$Se$_3$. This reaction can be described as the following oxidation
and reduction formulas in an electrochemical viewpoint:
\begin{equation}
x {\rm Cu} \rightarrow x {\rm Cu}^+  + xe^-
\end{equation}
\begin{equation}
{\rm Bi}_2{\rm Se}_3 + xe^- \rightarrow ({\rm Bi}_2{\rm Se}_3)^{x-}.
\end{equation}
The occurrence of electron transfer from Cu to Bi$_2$Se$_3$ implies that
the redox potential of Cu$^{+/0}$ is more negative than that of
(Bi$_2$Se$_3$)$^{0/x-}$. Since we are using a Cu wire to hold the sample
and to make the electrical contact, it is possible that this wire became
the source of an additional Cu$^+$ intercalation to the sample(s). This
reaction should not be included into the charge balance, and hence the
difference between the estimated and the theoretical Faraday number is
naturally expected.
This inference is supported by our observation that the sample mass
increased even without applying a current, though at a much lower
reaction rate.

It is most likely that the electrochemical intercalation is induced by
reduction of Bi$_2$Se$_3$ without reduction of Cu$^+$, yielding
(Cu$^{+}$)$_x$(Bi$_2$Se$_3$)$^{x-}$. This implies that doping of electrons
with the nominal fraction of $x$ should take place upon Cu$^+$
intercalation. The reaction at the counter electrode seems to be
oxidation of Cu metal rather than that of I$^{-}$ to I$_2$ or I$_{3}^{-}$,
because the former redox potential is more negative than the latter one
in an aqueous medium.

It is important to note that the as-intercalated samples \textit{do not}
superconduct yet. It turned out that the samples have to be annealed in
order to establish superconductivity. For this purpose, the samples were
sealed under vacuum in quartz glass tubes and put into a muffle furnace.
For the determination of the optimum annealing temperature, one sample
with $x$ = 0.31 was annealed subsequently for 2 h at different
temperatures starting at $500^{\circ}$C. After each annealing run, the
diamagnetic response was checked, and the result is shown in the inset
of Fig.~\ref{setup}(b). A trace of diamagnetism indicating the
appearance of a superconducting phase was detected after annealing at
530$^{\circ}$C, and the annealing temperature of 560$^{\circ}$C was
found to yield the largest shielding fraction. When the sample was
annealed at higher temperatures than 560$^{\circ}$C, the shielding
fraction was found to be reduced irreversibly. (The onset $T_c$ was 
essentially independent of the annealing temperature.) 
Therefore, the samples
used for determining the phase diagram presented here were treated in
the following way: They were heated up to 540$^{\circ}$C in 1 h, and
then the temperature was gradually increased to 560$^{\circ}$C in 40 min
to avoid any overheating; the samples were kept at 560$^{\circ}$C for 2
h and eventually quenched by dropping the quartz glass tubes into cold
water.\cite{time} We will come back later to the question of what is
happening during the annealing process to activate the
superconductivity.

To confirm the accuracy of the $x$ values determined by the mass change
(and also to make sure that the Cu content does not change appreciably
during the annealing process), we measured the $x$ values of eight of
our samples in the post-annealed state by using the ICP-AES analysis
(the sample mass was between 15 to 38 mg). For this destructive
analysis, the whole sample was dissolved in nitric acid HNO$_3$. The Cu
concentrations obtained from the ICP-AES analyses for the eight
samples agreed with those obtained from the mass change within
$\pm$0.014,\cite{ICP} giving confidence in the $x$ values reported in this
paper.

\begin{figure}[t]
\includegraphics[width=8.5cm,clip]{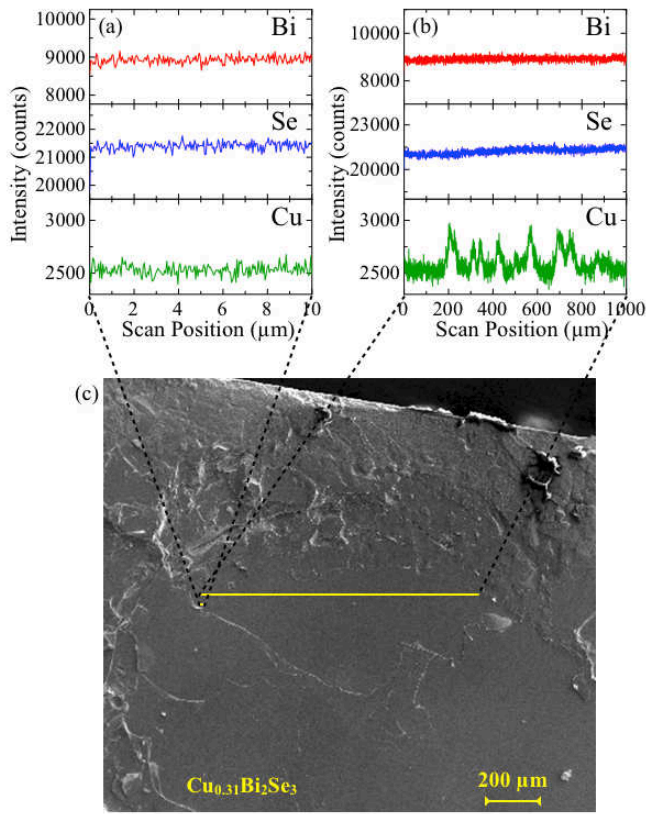}
\caption{(color online) 
EPMA analyses of the cleaved surface of a sample with $x=0.31$. Panels
(a) and (b) summarize scans along two lines of 10 $\mu$m and 1 mm
lengths, respectively, sketched in panel (c). 
One can see that the local Cu concentration
is uniform on the $\mu$m scale, but it is inhomogeneous on the sub-mm
scale.}
\label{EPMA}
\end{figure}

In addition to the ICP-AES analysis, we have employed the electron-probe
micro-analyzer (EPMA) to check the distribution of Cu within the sample
after the annealing. For this analysis, one sample with $x$ = 0.31 was
cleaved and subsequently scanned over distances of 10 $\mu$m and 1 mm,
as sketched in Fig.~\ref{EPMA}(c). The intensity of the characteristic
X-ray of each element is plotted vs the scan position for the two
different scan lengths in Figs. 2(a) and 2(b). One can see that the
distributions of Bi and Se are essentially homogeneous on any length scale,
as evidenced by the constant intensity of the respective characteristic
X-ray. On the other hand, the distribution of Cu shows a variation of up
to $\sim$20\% on the sub-mm length scale [Fig. 2(b)], although on the
10-$\mu$m length scale the Cu distribution is usually homogeneous [Fig.
2(a)]. Since the averaged Cu concentration of this sample determined
from the mass change was 0.31, the spatial variation of $\sim$20\%
corresponds to the variation in $x$ of $\sim$0.06.

\section{Sample Characterizations}

\begin{figure}[t]
\includegraphics[width=8.5cm,clip]{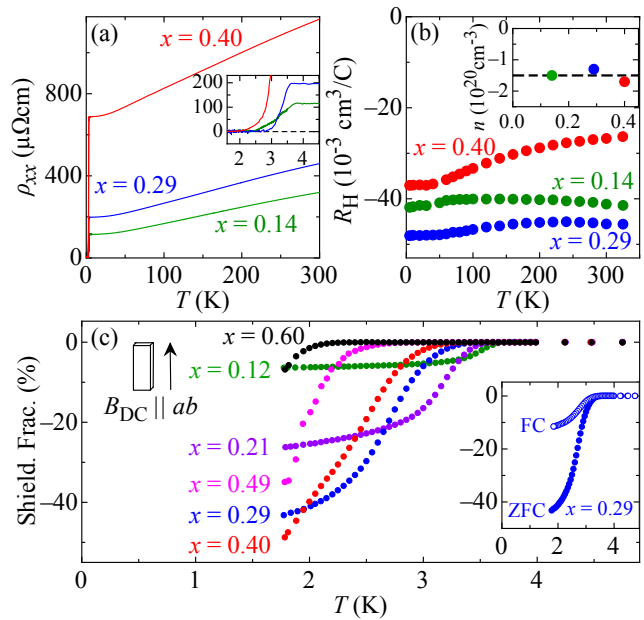}
\caption{(color online)
(a) Temperature-dependent resistivity of Cu$_{x}$Bi$_2$Se$_3$ 
for $x=0.14$, 0.29, and 0.40. The inset gives an enlargement around the 
superconducting transition. (b) Temperature dependence of the Hall 
coefficient $R_{\rm H}$ and the $x$ dependence of the charge-carrier 
concentration $n$ (inset). (c) Temperature-dependence of 
the superconducting shielding fraction in $B = 0.2$ mT after ZFC for 
various Cu concentrations $0.12 \leq x\leq 0.60$. For $x=0.29$ the 
ZFC and FC data are exemplarily shown in the inset.}
\label{rhochiplot}
\end{figure}

The superconducting samples were characterized by measuring the dc
magnetization $M$ and transport properties. To take the magnetization
data, a commercial SQUID magnetometer (Quantum Design, MPMS) was used
with the magnetic field applied parallel to the $ab$ plane. 
The samples were cooled down in zero-magnetic field (zero-field
cooled, ZFC) to the lowest accessible temperature $T = 1.8$ K; then a
small dc field of $B = 0.2$ mT was applied and the magnetization was
measured upon increasing temperature. After passing through $T_{\rm c}$,
defined as the onset of the drop in the $M(T)$ curves, data were again
taken upon decreasing the temperature back to 1.8 K (field cooled, FC).
The superconducting shielding fraction of a sample was estimated from
its magnetic moment at $T=1.8$ K after ZFC. The resistivity $\rho_{xx}$
and the Hall coefficient $R_H$ were measured by a standard six-probe
technique, where the
electrical current was applied in the $ab$ plane.

With our electrochemical intercalation technique, we have successfully
synthesized samples which superconduct above 1.8 K for Cu concentrations
of $0.09 \leq x \leq 0.64$. Figure~\ref{rhochiplot}(a) 
shows the resistivity data
of Cu$_{x}$Bi$_2$Se$_3$ for three selected Cu concentrations $x=0.14$,
0.29, and 0.40 measured in zero field. All those samples exhibit a
metallic temperature dependence above $T_{\rm c}$ and show zero
resistance below $T_{\rm c}$, see the expanded view in the inset of
Fig.~\ref{rhochiplot}(a). With increasing Cu concentration, the absolute value
of $\rho_{xx}$ increases; especially, between $x=0.29$ and 0.40 a strong
rise in the absolute value is observed, which implies that a high Cu
concentration enhances the disorder in the samples. Figure~\ref{rhochiplot}(b)
summarizes the temperature dependences of $R_H$ for the three samples, 
which are generally weak.
The inset shows the $x$-dependence of the charge-carrier concentration
$n$ determined from the low-temperature value of $R_H$. It is striking that, 
despite the factor of three difference in the Cu
concentration, the change in $n$ is very small: $n$ =
1.5$\times$10$^{20}$, 1.3$\times$10$^{20}$, and 1.7$\times$10$^{20}$
cm$^{-3}$ for $x$ = 0.14, 0.29, and 0.40, respectively. 
Furthermore, those values correspond to only $\sim$2\% of electron doping.
This is totally inconsistent with the expectation that there should be
electron carriers with the nominal fraction of $x$ in Cu$_{x}$Bi$_2$Se$_3$. 
Therefore, there must be some
side reaction taking place in Cu$_{x}$Bi$_2$Se$_3$ to significantly
reduce the actual electron carriers. In this regard, a
similar problem was previously noted in a
study of Cu intercalation into Bi$_2$Te$_3$, and it was proposed that
Cu$^+$ reacts with the matrix to form a four-layer lamellar structure
Cu-Te-Bi-Te that annihilates electrons.\cite{bludska04a} If a similar
reaction occurs in Cu$_{x}$Bi$_2$Se$_3$, it would be
\begin{equation}
{\rm Cu}^{+} + 2{\rm Bi}_2{\rm Se}_3 + e^- \rightarrow 
[{\rm CuBiSe}_2][{\rm Bi}_3{\rm Se}_4^-] + h.
\end{equation}
The formation of this type of structural defects can indeed explain the
small $n$ values observed in Cu$_{x}$Bi$_2$Se$_3$.

Figure~\ref{rhochiplot}(c) summarizes the temperature dependences of the
shielding fraction in samples with various $x$ values. The onset
temperatures of superconductivity in the magnetization data are slightly
lower than the onset of the resistivity transition in respective
samples, which is usual for disordered superconductors. The right inset
of Fig.~\ref{rhochiplot}(c) shows the ZFC and FC data for $x=0.29$, and other
samples exhibited similar differences between the two. We note that
samples with $x\geq 0.70$ were also prepared, but they did not show any
superconducting transition above 1.8 K; however, it is possible that
they exhibit superconductivity at lower temperatures.

\section{Phase Diagram and Discussions}

\begin{figure}[t]
\includegraphics[width=8.5cm,clip]{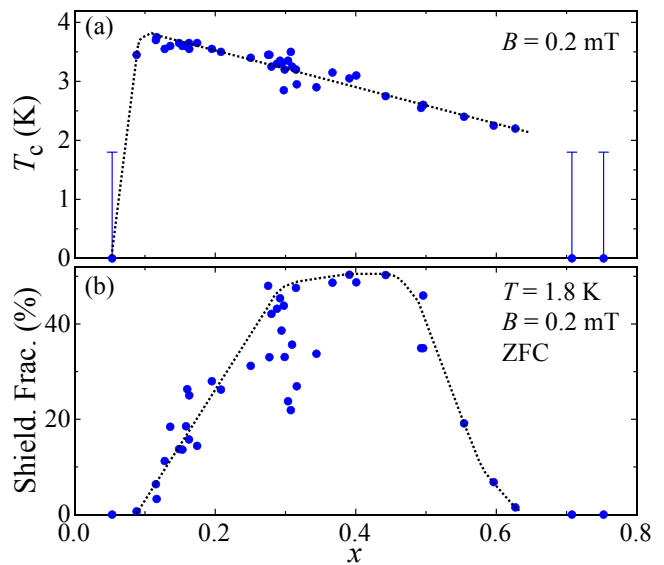}
\caption{(color online)
Cu concentration $x$ dependence of (a) the critical temperature and 
(b) the shielding fraction at 1.8 K. These $T_{\rm c}$ data were extracted 
from the onset of superconductivity in magnetization measurements, the 
shielding-fraction data from the ZFC magnetization at $T=1.8$ K, 
see Fig.~\ref{rhochiplot}(c).}
\label{Tcxplot}
\end{figure}
Figure~\ref{Tcxplot}(a) shows the phase diagram of $T_{\rm c}$ vs $x$ based on
our measurements of more than 40 samples, and Fig.~\ref{Tcxplot}(b) summarizes
the shielding fraction of those samples at 1.8 K. The lowest $x$ value
at which we found superconductivity was 0.09, where $T_{\rm c}$ was
already 3.45 K; however, in spite of this relatively high $T_{\rm c}$,
the superconducting signal was very weak with a shielding fraction of
only around 1\%. For slightly larger $x$ values around 0.12--0.15 the
maximum values of $T_{\rm c}$ are found, but the shielding fractions were
less than 20\%, which is in agreement with the earlier report by Hor
\textit{et al.}\cite{hor10a} Intriguingly, $T_{\rm c}$ gradually
decreases with increasing $x$ down to $T_{\rm c}\approx 2.2$ K for $x$ =
0.64, which is the highest Cu concentration for which we could clearly
confirm the superconductivity.

Although $T_{\rm c}$ seems to be quite robust and reproducible, the
shielding fraction exhibits a scattering among different samples even
with similar $x$ values, as can be seen in Fig.~\ref{Tcxplot}(b).
Nevertheless, there is a clear trend that samples with maximal shielding
fractions are obtained only for $0.3 \le x \le 0.5$ and the shielding
fraction is reduced systematically (almost linearly) as $x$ becomes
smaller than $x \simeq 0.3$. This trend, 
together with the EPMA result that the actual Cu concentration
varies on a sub-mm scale, suggests that Cu$_{x}$Bi$_2$Se$_3$
has a tendency to phase-segregate and that there is a spontaneous 
formation of small
islands in which the local Cu concentration is essentially the optimum
value for superconductivity. In view of the fact that the $T_c$ of this
material is uncharacteristically high for a low-carrier-density
superconductor \cite{Cohen} with $n \sim 10^{20}$ cm$^{-3}$, the present
observation seems to imply that the inhomogeneous nature of the
superconductivity may partly be responsible for the anomalously high
$T_c$.\cite{Kivelson} Furthermore, the unexpected drop of the shielding
fraction for $x > 0.5$ suggests that the inhomogeneity may be necessary
for the occurrence of superconductivity in this material.

If Cu$_{x}$Bi$_2$Se$_3$ indeed has a tendency toward phase segregation
to spontaneously form small superconducting islands, it naturally explains
why a high-temperature annealing is necessary for activating the
superconductivity: It allows Cu atoms to move and promotes the formation
of superconducting islands. If annealed at too high a temperature, the
material seems to melt partly, the crystal structure degrades and
eventually the superconductivity is destroyed irreversibly. Indeed, the
melting temperature of Cu$_{x}$Bi$_2$Se$_3$ appears to be systematically
reduced with $x$, and our test sample with $x$ = 1.4, which was
initially rectangular shaped, became sphere-like after annealed at
560$^{\circ}$C. (Note that the melting point of pure Bi$_2$Se$_3$ is
706$^{\circ}$C). Therefore, the optimal annealing temperature
corresponds to the one that maximally promotes the motion of Cu atoms
while avoiding degradation of the matrix due to partial melting.

\section{Summary}

In summary, we have synthesized superconducting Cu$_{x}$Bi$_2$Se$_3$ by
an electrochemical intercalation method, which yields samples with much
higher superconducting shielding fractions compared to those prepared by
the melt-growth technique previously employed. The superconductivity was
observed for $0.09 \le x \le 0.64$, which is much wider than previously
reported, and the transition temperature $T_c$ was found to show an
unexpected monotonic decrease with $x$. Also, the largest attainable
shielding fraction was found to be strongly dependent on $x$, and its
$x$ dependence suggests that Cu$_{x}$Bi$_2$Se$_3$ has a
tendency to phase-segregate and form small superconducting islands.

\begin{acknowledgments}
We acknowledge the technical support from the Comprehensive Analysis
Center, Institute of Scientific and Industrial Research, Osaka
University, for the ICP-AES and the EPMA analyses. This work was
supported by JSPS (NEXT Program), MEXT (Innovative Area ``Topological
Quantum Phenomena" KAKENHI 22103004), and AFOSR (AOARD 10-4103).
\end{acknowledgments}


\begin{thebibliography}{10}
\parskip-0.2ex plus0.05ex minus0.05ex

\bibitem{K1} 
L. Fu, C. L. Kane, and E. J. Mele, Phys. Rev. Lett. {\bf 98}, 106803 (2007).

\bibitem{MB}
J.E. Moore and L. Balents, Phys. Rev. B {\bf 75}, 121306(R) (2007).

\bibitem{Roy}
R. Roy, Phys. Rev. B {\bf 79}, 195322 (2009).

\bibitem{hasan10a}
M.Z.~Hasan and C.L.~Kane,
\newblock Rev.\ Mod.\ Phys. {\bf 82}, 3045 (2010).

\bibitem{moore10a}
J.E.~Moore,
\newblock Nature {\bf 464}, 194 (2010).

\bibitem{qi10d}
X.-L. Qi and S.-C.~Zhang,
\newblock arXiv:1008.2026v1 (2010).

\bibitem{K2}
L. Fu and C.L. Kane, Phys. Rev. B {\bf 76}, 045302 (2007).

\bibitem{SCZ4}
H.-J. Zhang, C.-X. Liu, X.-L. Qi, X. Dai, Z. Fang, and S.-C. Zhang, 
Nat. Phys. {\bf 5}, 438 (2009).

\bibitem{H1} 
D. Hsieh, D. Qian, L. Wray, Y. Xia, Y. S. Hor, R. J. Cava, and 
M. Z. Hasan, Nature {\bf 452}, 970 (2008).

\bibitem{Taskin}
A. A. Taskin and Y. Ando, Phys. Rev. B {\bf 80}, 085303 (2009). 

\bibitem{Matsuda}
A. Nishide, A. A. Taskin, Y. Takeichi, T. Okuda, A. Kakizaki, 
T. Hirahara, K. Nakatsuji, F. Komori, Y. Ando, and I. Matsuda, 
Phys. Rev. B. {\bf 81}, 041309(R) (2010).

\bibitem{Shen1}
Y. L. Chen, J. G. Analytis, J.-H. Chu, Z. K. Liu, S.-K. Mo, X. L. Qi, 
H. J. Zhang, D. H. Lu, X. Dai, Z. Fang, S. C. Zhang, I. R. Fisher, 
Z. Hussain, and Z.-X. Shen, Science {\bf 325}, 178 (2009). 

\bibitem{H4}
D. Hsieh, Y. Xia, D. Qian, L. Wray, F. Meier, J. H. Dil, J. Osterwalder, 
L. Patthey, A. V. Fedorov, H. Lin, A. Bansil, D. Grauer, Y. S. Hor, 
R. J. Cava, and M. Z. Hasan, Phys. Rev. Lett. {\bf 103}, 146401 (2009).

\bibitem{H5}
Y. Xia, D. Qian, D. Hsieh, L. Wray, A. Pal, H. Lin, A. Bansil, D. Grauer, 
Y. S. Hor, R. J. Cava, and M. Z. Hasan, Nat. Phys. {\bf 5}, 398 (2009).

\bibitem{Sato}
T. Sato, K. Segawa, H. Guo, K. Sugawara, S. Souma, T. Takahashi, and Y. Ando, 
Phys. Rev. Lett. {\bf 105}, 136802 (2010). 

\bibitem{Kuroda}
K. Kuroda, M. Ye, A. Kimura, S. V. Eremeev, E. E. Krasovskii, E. V. Chulkov,
Y. Ueda, K. Miyamoto, T. Okuda, K. Shimada, H. Namatame, and M. Taniguchi,
Phys. Rev. Lett. {\bf 105}, 146801 (2010). 

\bibitem{Shen2}
Y. L. Chen, Z. K. Liu, J. G. Analytis, J.-H. Chu, H. J. Zhang, B. H. Yan,
S.-K. Mo, R. G. Moore, D. H. Lu, I. R. Fisher, S.-C. Zhang, Z. Hussain,
and Z.-X. Shen, Phys. Rev. Lett. {\bf 105}, 266401 (2010). 

\bibitem{BTS_Rapid} 
Z. Ren, A.A. Taskin, S. Sasaki, K. Segawa, and Y. Ando, 
Phys. Rev. B {\bf 82}, 241306(R) (2010). 

\bibitem{BTS_Hasan}
S.Y. Xu, L.A. Wray, Y. Xia, R. Shankar, A. Petersen, A. Fedorov, H. Lin, A. Bansil, 
Y.S. Hor, D. Grauer, R.J. Cava, and M.Z. Hasan,
arXiv:1007.5111v1.

\bibitem{Xiong}
J. Xiong, A.C. Petersen, Dongxia Qu, R. J. Cava, and N. P. Ong, 
arXiv:1101.1315.

\bibitem{BSTS}
A.A. Taskin, Z. Ren, S. Sasaki, K. Segawa, and Y. Ando, 
Phys. Rev. Lett. {\bf 107}, 016801 (2011). 

\bibitem{fu08a}
L.~Fu and C.L.~Kane,
\newblock Phys.\ Rev.\ Lett. {\bf 100}, 096407 (2008).

\bibitem{schnyder08a}
A.P. Schnyder, S.~Ryu, A.~Furusaki, and A.W.W. Ludwig,
\newblock Phys.\ Rev.\ B {\bf 78}, 195125 (2008).

\bibitem{qi09b}
X.-L. Qi, T.L. Hughes, S.~Raghu, and S.-C. Zhang,
\newblock Phys.\ Rev.\ Lett. {\bf 102}, 187001 (2009).

\bibitem{qi10b}
X.-L.~Qi, T.L.~Hughes, and S.-C.~Zhang,
\newblock Phys.\ Rev.\ B {\bf 81}, 134508 (2010).

\bibitem{linder10b}
J.~Linder, Y.~Tanaka, T.~Yokoyama, A.~Sudb\o, and N.~Nagaosa,
\newblock Phys.\ Rev.\ Lett. {\bf 104}, 067001 (2010).

\bibitem{sato10a}
M.~Sato,
\newblock Phys.\ Rev.\ B {\bf 81}, 220504(R) (2010).

\bibitem{fu10a}
L.~Fu and E.~Berg,
\newblock Phys.\ Rev.\ Lett. {\bf 105}, 097001 (2010).

\bibitem{hor10a}
Y.S. Hor, A.J. Williams, J.G. Checkelsky, P.~Roushan, J.~Seo, Q.~Xu, H.W.
  Zandbergen, A.~Yazdani, N.P. Ong, and R.J. Cava,
\newblock Phys.\ Rev.\ Lett. {\bf 104}, 057001 (2010).

\bibitem{wray10a}
L.A. Wray, S.-Y. Xu, Y.~Xia, Y.S. Hor, D.~Qian, A.V. Fedorov, H.~Lin,
  A.~Bansil, R.J. Cava, and M.Z. Hasan,
\newblock Nature Physics {\bf 6}, 855 (2010).

\bibitem{kriener11a}
M.~Kriener, K.~Segawa, Z.~Ren, S.~Sasaki, and Y.~Ando,
\newblock Phys.\ Rev.\ Lett. {\bf 106}, 127004 (2011).

\bibitem{caywood70a}
L.P. Caywood and G.R. Miller,
\newblock Phys.\ Rev.\ B {\bf 2}, 3209 (1970).

\bibitem{vasko74a}
A.~Va\v{s}ko, L.~Tich\'y, J.~Hor\'ak, and J.~Weissenstein,
\newblock Appl.\ Phys. {\bf 5}, 217 (1974).

\bibitem{accuracy}
For example, the wavelength dispersive X-ray spectroscopy (WDS)
analysis, which is a non-destructive method and is often employed in the
chemical content analysis, is sensitive to relative differences, but it is
not good at giving absolute numbers.

\bibitem{time}
As for the optimization of the annealing time, we have annealed several
samples that were already annealed at 560$^{\circ}$C for 2 h for various
additional durations of time at the same temperature; annealing for a
few additional hours did not improve the shielding fraction, and a
long-term annealing (more than a day) usually resulted in a decrease in
the shielding fraction. Separately, we found that an annealing for less
than 1 h did not establish superconductivity. We therefore settled on
the 2-h annealing procedure to minimize the sample preparation time.

\bibitem{ICP}
The detailed results of the ICP-AES analyses are the following: the 
$x$ values of the eight samples determined from the mass change were
0.117, 0.128, 0.266, 0.286, 0.293, 0.295, 0.308, and 0.318;
for these samples, the $x$ values determined by the ICP-AES analyses were 
0.114, 0.128, 0.279, 0.300, 0.299, 0.282, 0.303, and 0.319, respectively; 
therefore, the differences between the two were 
0.003, 0.000, $-0.013$, $-0.014$, $-0.006$, 0.013, 0.005, and $-0.001$.

\bibitem{bludska04a}
J.~Bludsk\'a, S.~Karamazov, J.~Navr\'atil, I.~Jakubec, and J.~Hor\'ak,
\newblock Solid-State Ionics {\bf 171}, 251 (2004).

\bibitem{Cohen}
M.L. Cohen, in {\it Superconductivity}, ed. by R.D. Parks, Vol. 1 
(Marcel Dekker, 1969), Chap. 12.

\bibitem{Kivelson}
I. Martin, D. Podolsky, and S.A. Kivelson,
Phys. Rev. B {\bf 72}, 060502(R) (2005). 


\end{thebibliography}
\end{document}